\documentstyle[graphicx]{mn}
\newcommand{\ls}
 {\mathrel{\hbox{\rlap{\hbox{\lower4pt\hbox{$\sim$}}}\hbox{$<$}}}}
\newcommand{\gs}
 {\mathrel{\hbox{\rlap{\hbox{\lower4pt\hbox{$\sim$}}}\hbox{$>$}}}}
\newcommand{\degg}{\hbox{$^\circ$}}

\newcommand{\arcs}{\hbox{$^{\prime\prime}$}}
\newcommand{\et}{et al.\ }
\newcommand{\ginga}{{\it Ginga}}
\newcommand{\rosat}{{\it ROSAT}}

\newcommand{\asca}{{\it ASCA}}
\newcommand{\xte}{{\it RXTE}}

\newcommand{\erg}{erg\,s$^{-1}$}
\def\la{\mathrel{\hbox{\rlap{\hbox{\lower4pt\hbox{$\sim$}}}{\raise2pt\hbox{$<$}}
}}}
\def\ga{\mathrel{\hbox{\rlap{\hbox{\lower4pt\hbox{$\sim$}}}{\raise2pt\hbox{$>$}}
}}}

\title{PDS 456: an Extreme Accretion Rate Quasar?}

\author[Reeves {\it et al.}]
{J.N. Reeves$^1$, P.T. O'Brien$^1$, S. Vaughan$^1$, D. Law-Green$^1$, 
M. Ward$^1$, 
\and
C. Simpson$^2$, K.A. Pounds$^1$, R. Edelson$^{1,3}$ \\
$^1$X-Ray Astronomy Group; Department of Physics and Astronomy;
Leicester University; Leicester LE1 7RH; U.K.\\
$^2$Subaru Telescope, National Astronomical Observatory of Japan, 650 N.
A`oh\={o}k\={u} Place, Hilo, HI 96720, U.S.A.\\ 
$^3$Department of Physics and Astronomy; University of California,
Los Angeles; Los Angeles, CA 90095-1562; U.S.A.}

\begin{document}
\maketitle
\begin{abstract}

We present quasi-simultaneous \asca\ and \xte\ observations of
the most luminous known AGN in the local (z\ $<0.3$) universe, the recently
discovered quasar
PDS~456. Multiwavelength observations have been conducted which show 
that PDS~456 has a bolometric luminosity of $\sim10^{47}$ \erg\,
peaking in the UV part of the spectrum. In the X-ray band the 2-10~keV (rest-frame) luminosity is 10$^{45}$~erg~s$^{-1}$. 
The broad-band X-ray spectrum obtained with
\asca\ and \xte\ contains considerable complexity. The most
striking feature observed is a very deep, ionised iron K edge,
observed at 8.7 keV in the quasar rest-frame. We find that these
features are consistent with reprocessing from highly ionised matter,
probably the inner accretion disk. 
PDS~456 appeared to show a strong (factor of $\sim$2.1) outburst in just
$\sim$17~ksec, although non-intrinsic sources cannot be completely
ruled out. If confirmed, this would be an unusual event for 
such a high-luminosity source, 
with a light-crossing-time corresponding to $\sim2R_S$. 
The implication would be that flaring occurs within the very
central regions, or else that PDS~456 is a `super-Eddington'
or relativistically beamed system. 
Overall we conclude on the basis of the extreme blue/UV luminosity,
the rapid X-ray variability and from the imprint of highly ionised
material on the X-ray spectrum, that PDS~456 is a quasar with an
unusually high accretion rate.

\end{abstract}


\section{Introduction}

Understanding the origin of the luminous, ionising continuum from Active
Galactic Nuclei (AGN) is one of the prime goals of AGN research. Despite
the general acceptance of the standard AGN paradigm --- that the
continuum (UV) emission originates in an 
accretion disk around a super-massive
black hole --- the details remain unclear. In the X-ray band, there is
a hard power-law
component that dominates above 2~keV in the well studied
Seyfert~1 galaxies. This hard X-ray continuum
may originate from a hot corona above the surface of the accretion disk, in
which optical/UV photons from the disk are Comptonised to X-ray energies.
These X-rays then illuminate the disk, being either `reflected'
towards the observer or thermalised back into
optical/UV emission. In the X-ray spectral band, 
evidence for this disk reflection component 
is seen in the form of a fluorescent Fe K$\alpha$ line 
at 6.4~keV, an iron K edge at $>7$~keV and a
Compton `hard tail'. All these features have been observed in
Seyfert~1 galaxies (e.g. Pounds \et 1990, Nandra \& Pounds 1994).  

%To determine whether the Seyfert~1 X-ray spectral features are evident
%in high-luminosity AGN, particularly radio-quiet quasars,
%it is essential to observe those
%objects which are the most extreme in luminosity and/or continuum shape.

The subject of this paper is the recently discovered {\it radio-quiet} 
quasar PDS~456 (Torres \et 1997, Simpson \et 1999), which is {\it the} most 
luminous known object in the local Universe (z\ $<0.3$); indeed the bolometric
luminosity of PDS~456 exceeds that of the well-known radio-loud quasar
3C 273. However, unlike 3C~273, PDS~456 is radio-quiet, and thus
presumably not jet dominated. Therefore,
PDS~456 provides a unique opportunity to study in detail the hard and soft
X-ray components in a high luminosity, radio-quiet AGN, and in
particular to test models of reflection expected from the putative
accretion disk.

In the next section we describe the multi-wavelength
properties of PDS~456. The main subject of this paper is then covered; 
the X-ray emission from PDS~456, using observations with \asca\ and \xte. 
Values of $ H_0 = 50 $~km\,s$^{-1}$\,Mpc$^{-1}$ and $ q_0 = 0.5 $ have been
assumed throughout and all fit parameters are given in the quasar rest-frame.
Note that errors are in this paper are quoted at
the 90\% confidence level (e.g. $\Delta \chi^{2}=2.7$ for one 
interesting parameter). 

\section{Multiwavelength properties of PDS~456}

%The luminous quasar PDS~456 was recently identified by Torres {\it et al.}
%(1997) in a Galactic search for young stellar objects. 
%Since, we have conducted an extensive multi-wavelength campaign to
%observe the properties of PDS 456 across the whole electromagnetic 
%spectrum. 
%the optical and near infra-red properties of PDS are discussed 
%in Simpson \et 1999.

At $z=0.184$, PDS~456 has a de-reddened, absolute blue magnitude
M$_{B}\approx -27$, making it more luminous than the radio-loud quasar
3C~273 ($z=0.158$, M$_{B}\approx -26$).
The Galactic absorption column towards PDS~456 is $2.4\times10^{21}$~cm$^{-2}$
(Stark \et 1992), with an extinction of A$_{V}=1.4$ in the optical. 
We have performed a recent 
VLA observation; the total detected 6~cm flux of 8~mJy
confirms that PDS~456 is radio-quiet 
(R$_{L}$ = log[F$_{6cm}$/F$_{B}$]~$\approx$~-0.74). 
PDS~456 has broad Balmer and Paschen lines, 
strong Fe \textsc{ii}, a hard (de-reddened)
optical continuum ($f_{\nu} \propto \nu^{-0.1\pm0.1}$), and one of the
strongest `big blue bumps' of any AGN (Simpson \et 1999). 
PDS~456 is coincident (to $<$ 10\arcs) 
with the \rosat\ soft X-ray source RXS~J172819.3-141600
(Voges {\it et al.} 1999); the count rate is $0.30\pm0.02$
ct/s corresponding to L$_{X}\sim10^{45}$ \erg. 
%Given the high positional accuracy of the \rosat\ X-ray source, 
%the X-rays are likely to originate directly from the PDS~456.
The overall spectral energy distribution of PDS~456 (figure 1), 
obtained from all our current data on this quasar 
(including recent VLA and ISO data, listed in table 1), 
gives a bolometric luminosity of
$\sim10^{47}$\erg, a value more typical of QSOs at $z\sim3$. 

\vspace*{-0.5cm}

\begin{figure}
\begin{center}
\rotatebox{-90}{\includegraphics[width=5.5cm]{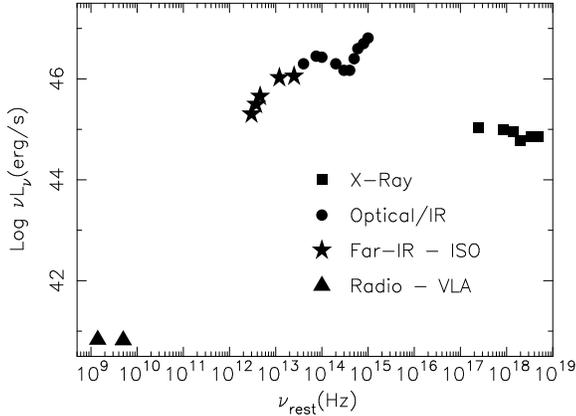}}
\end{center}
\caption{The radio to X-ray SED of the quasar PDS~456. X-ray
(ROSAT, ASCA, RXTE), infra-red (ISO) and radio (VLA) data are plotted; the
dereddened 
optical/near-IR data has been taken from Simpson \et (1999). It is seen
that the bolometric luminosity of PDS~456 
approaches $10^{47}$ \erg.
}
\end{figure}

\section{X-ray observations and Data reduction}

\subsection{ASCA observations}

PDS~456 was observed by \asca\ on 7--8 March 1998 with a total 
duration of $\sim$140~ks.
Standard screening criteria were applied to the \asca\ data
(e.g. see the {\it ASCA Data Reduction Guide}), giving a
total `good' exposure time of $\sim40$~ksec per detector.
Counts were extracted using a 4\arcmin\ circular aperture
centered on the source, and the background was estimated
using source-free regions at similar off-axis angles. 
%The result of
%this screening and background subtraction yielded 
%4225 total source counts for the SIS0 and 4707 counts for GIS3.  
%The pulse height spectra were then binned to give at least 20 counts per
%spectral channel. 
%In addition source and background light curves 
%were extracted in 128 seconds bins,
%from both SIS detectors, in three bands: full (0.5--10~keV), soft
%(0.5--2~keV) and hard (2--10~keV).
%The two SIS background subtracted light curves were combined to
%increase the signal/noise. The resultant light curves have then been
%binned into orbital bins (see later) for subsequent analysis.

\begin{table}
\centering
\caption{ISO (ISOPHOT) and VLA fluxes.}
\begin{tabular}{lccc}\hline                 
\multicolumn{2}{c}{ISO Data} & \multicolumn{2}{c}{VLA Data} \\
$\lambda$($\mu$m) & Flux(Jy) & $\nu$(Hz) & Flux(mJy) \\\hline
12 & $0.28\pm0.03$ & 1.4 & 30.4 \\
25 & $0.55\pm0.12$ & 4.85 & 8.23 \\
50 & $0.61\pm0.11$ \\
80 & $0.51\pm0.11$ \\
100 & $0.42\pm0.14$ \\\hline 
\end{tabular}
\end{table}

\subsection{RXTE observations}

PDS~456 was observed by \xte\ on 7-10 March 1998 with a total 
duration of 226~ks.
We extracted \textsc{standard-2} data from the PCU 
(Proportional Counter Unit) array using the
\textsc{rex} reduction script supplied by NASA/GSFC.
Data have been extracted from Layer 1 (L1) only (the first and most sensitive
layer). 
%using PCUs 0, 1 and 2 for spectroscopy and PCUs 0, 1, 2 and 3
%for timing analysis.
Poor quality data were excluded on the basis of the following
selection criteria:
(i) the satellite is out of the South Atlantic Anomaly (SAA); 
(ii) Earth elevation angle $\geq$~10\degg; (iii) offset from the optical
position of PDS~456 is $\leq$~0.02\degg; and finally (iv)
\textsc{electron-0}$\leq$~0.1. This last criterion removes data
with high anti-coincidence rate in the PCUs. The total `good'
exposure time selected after this screening  was 92~ksec.
The background was estimated using the \textsc{l7-240} model.
%(See Edelson \& Nandra (1999) for a discussion of the effectiveness 
%of this model.) 
%The pulse height spectrum from the PCU
%array was also binned to give at least 20 counts per spectral channel, and
%an appropriate response matrix was generated using the
%\textsc{pcarsp} script. 
All \xte\ spectral data below 2.5~keV and above 18~keV have been ignored.

\section{ Spectral analysis }

%The background subtracted spectra from both satellites were fitted
%using the \textsc{xspec v10.0} software package; errors are quoted at
%the 90\% confidence level (e.g. $\Delta \chi^{2}=2.7$ for one 
%interesting parameter). 
%Initially the spectra in the overlapping 2.5-10 keV band for both
%\asca\ and \xte\ were fitted separately. However, the results in this
%band were found to be consistent, and no significant evidence was
%found for spectral variability within the \xte\ observation. 
%Thus we proceed to fit
%the time-averaged \asca\ and \xte\ spectra simultaneously, 
%allowing the relative normalisation of the instruments to vary. The
%Galactic absorption column towards PDS~456 is $2.4\times10^{21}$~cm$^{-2}$,
%from radio survey data (e.g. Stark \et 1992); this is 
%included as a {\it fixed} component in all the spectral fits. 

%(Note that this value of $N_{\rm H}$ is consistent with the reddening of
%A$_{V}\approx1.5$ derived independently by Torres \et 1997 or 
%Simpson \et 1999.)

Initially the hard X-ray (\xte +\asca) data in the 2.5-18 keV range
were analysed, using background subtracted spectra and by allowing the 
relative normalisations of all 5 instruments to vary.
The pulse height spectra from each detector 
were binned to give at least 20 counts per spectral channel.  
Firstly, a power-law plus fixed Galactic absorption (model 1) 
gives a poor fit ($\chi_{\nu}^{2} = 1.62$) to the data (see Table~2
for a summary of the model fits). Clear deviations in the data--model 
residuals are present around the iron K-shell regime, both in the
\asca\, and particularly the \xte\ data (see figure 2). We therefore
added an iron K edge and line to the fits, to
model the effects of Compton reflection off the accretion
disk, as observed in many Seyfert 1s
(e.g. Pounds \et 1990). We first considered
a neutral (but narrow; $\sigma=0.01$~keV) 
iron line and edge (model 2) fixed at 6.4 and 7.1 keV
respectively. Although there is some improvement
in the model fits, the overall fit remained unacceptable. 
A further improvement was obtained using a broad Fe line (model 3),
although the fit was still inadequate as an edge was apparent in
the data residuals at $>7$~keV. The derived line strength
is rather high (EW $\sim900$~eV); the line is also unusually broad
($\sigma\sim1.4$~keV) and redshifted (E $\sim5.6$~keV). 

\begin{figure}
\begin{center}
\rotatebox{-90}{\includegraphics[width=5.5cm]{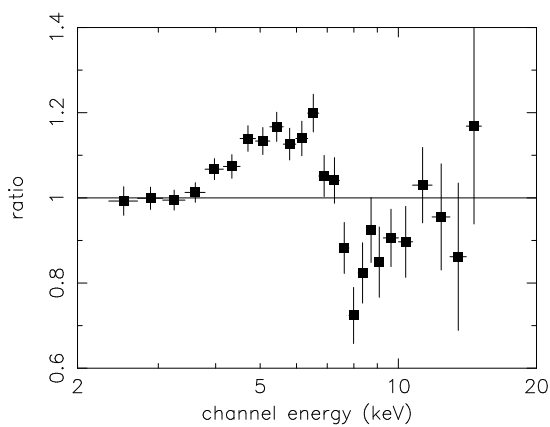}}
\end{center}
\caption{ The data/model ratio residuals from a simple power-law fit 
fit (model 1) to the \xte\ data of PDS~456. The effect of
the deep, ionised iron K edge is clearly seen in the residuals. }
\end{figure}

\begin{table*}
\centering
\caption{Results of simultaneous spectral fits to the \asca\ and \xte\ data
in the 2.5--18~keV range. 
Column 1 gives the models as defined in the text.
Columns 2--7 give the fit parameters: 
column 2 gives the power law photon index; 
column 3 the iron K edge energy (keV) or ionization parameter ($\xi$); 
column 4 gives the edge depth ($\tau$), reflection strength (R$=\Omega/2\pi$) or $N_{H}$ (units 10$^{22}$~cm$^{-2}$). 
Columns 5, 6 and 7 give the iron K line energy (keV), the intrinsic width
$\sigma$ (in keV) and equivalent width (in eV) respectively. 
Column 8 gives the best-fit $\chi^{2}$ over number of degrees of freedom.
$^{f}$ indicates the parameter is fixed.}
 
\begin{tabular}{@{}lccccccc@{}}                 
Model & $\Gamma$ & E$_{edge}$ or $\xi$ & $\tau$, R or $N_{H}$ & E$_{line}$ &
$\sigma$ & EW & $\chi^{2}$/dof \\   
 (1)    &  (2)     & (3)  & (4) & (5) & (6) & (7) & (8)  \\
\hline
1. PL & 2.38$\pm$0.03 & -- & -- & -- & -- & -- & 343/212 \\
2. PL+LINE & 2.42$\pm$0.04 & 7.1$^{f}$ & $<0.10$ & 6.4$^{f}$
& 0.01$^{f}$ & 178$\pm$49 & 309/210 \\
3. PL+BROAD-LINE & 2.44$\pm$0.05 & -- & -- & 5.6$^{+0.5}_{-0.6}$
& 1.4$^{+0.5}_{-0.4}$ & 900$^{+650}_{-300}$ & 230/209 \\
4. PL+EDGE+LINE & 2.12$\pm$0.04 & 8.70$\pm$0.15 & 0.76$\pm$0.12 &
6.7$^{f}$ & 0.01$^{f}$ & $<77$ & 211/209 \\
5. PL+EDGE+B-LINE & 2.22$\pm$0.08 & 8.7$\pm0.3$ & 0.53$\pm$0.18 &
6.1$\pm$0.5 & 1.2$^{+0.8}_{-0.7}$ & 340$^{+410}_{-200}$ & 200/207 \\ 
%4. PL+PEXRAV+LINE & 2.62$\pm$0.03 & -- & 1.0$^{f}$ & 6.4$^{f}$ &
%150$\pm$47 & 317/211 \\
6. PL+PEXRIV+LINE & 2.54$\pm$0.07 & 6400$^{+2650}_{-2350}$ &
0.92$^{+0.39}_{-0.27}$ & 6.7$^{f}$ & 0.01$^{f}$ & $<30$ & 207/209 \\
7. PL+ABSORI+LINE & 2.39$\pm$0.08 & 1.5$^{+0.6}_{-0.4}\ \times\
10^{4}$ & 47$^{+16}_{-11}$ & 6.7$^{f}$ & 0.01$^{f}$ & 
$<35$ & 199/209 \\
\hline
\end{tabular}
\end{table*}

We next tried a fit to highly ionised iron (model 4), the 
(narrow) line energy was fixed at 6.7 keV (corresponding to He-like
Fe) and the edge energy was left as a free parameter. In
this case an acceptable fit was found to the data 
($\chi_{\nu}^{2}\approx1$); with a deep and highly ionised iron K edge
at 8.7 keV with $\tau=0.76\pm0.12$. A {\it narrow} 
iron emission line
was no longer required in the fit (EW$<77$ eV). The energy of the edge
is consistent with iron of ionisation from Fe~\textsc{xxiii} --
Fe~\textsc{xxv}. Figure 3 clearly illustrates the high ionisation
state of the reprocessing material. Note that we also tried fitting a 
smeared edge model (width 1 keV), as might be expected from the inner disk, 
although the fit obtained was very similar to above. 
Including a {\it broad} iron line in the
model (in addition to the edge, model 5) lead to a further improvement in the
spectral fit ($\Delta\chi^{2}\sim11$), at a significance level of
$\sim99\%$ confidence (F-test, 2 additional parameters). The line
equivalent width 
is $\sim350$~eV; the line is broad ($\sigma\sim1$~keV) and
possibly redshifted (E $\sim6$~keV). We note that the detection of the ionised
iron edge is robust, even after the inclusion of a broad line 
in the model fit.

\begin{figure}
\begin{center}
\rotatebox{-90}{\includegraphics[width=5.5cm]{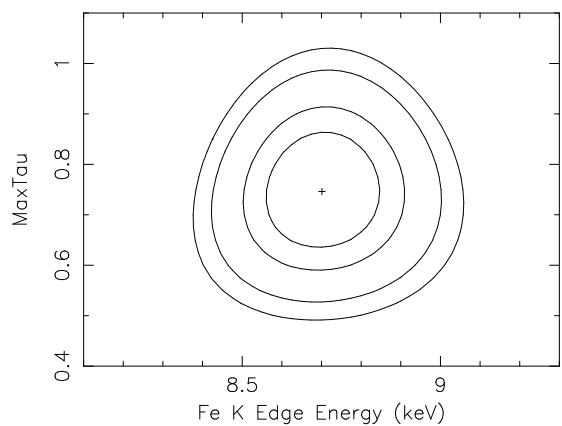}}
\end{center}
\caption{ Confidence contours of Fe K edge energy against depth, for
the \xte\ and \asca\ spectrum of PDS~456. The contours represent the 
68\%, 90\%, 99\%, 99.9\% significance levels respectively.
%The edge clearly originates from high ionisation material.
}
\end{figure}

Having established the presence of reprocessing by highly ionised
matter, we then attempted to model these
reprocessing features in terms of reflection of X-rays off the surface
layers of an ionised accretion disk. 
The \textsc{pexriv} model in XSPEC was used (Magdziarz \& Zdziarski
1995), assuming a
disk surface temperature of 10$^{6}$ K (e.g. Zycki \et 1994; Ross, Fabian
\& Young 1999) and a disk inclination angle of 30\degg. 
The disk ionisation parameter ($\xi$) and the strength
of the reflection component R (=$\Omega/2\pi$) 
were both left as free parameters in the
fit. (Note that $\xi=L/nr^{2}$, where $n$ is the number
density of the material at a distance $r$ from an ionising source of
luminosity $L$, where $L$ is defined from 5 eV to 20 keV.)
An ionised reflector provided a good fit to the hard X-ray spectrum
and the iron K edge (model 6, $\chi_{\nu}^{2}\approx1$), 
giving a value of R close to 1 and a high ionisation
($\xi=6400$~erg\ cm\ s$^{-1}$). 
Although a narrow iron K line is not required in this fit,
the presence of a broad line was not ruled out (EW $\la250$~eV).

We also attempt to model the hard X-ray spectrum
of PDS~456 with a power-law modified by a warm line-of-sight absorber 
(model 7), using the \textsc{absori} routine in XSPEC 
(note we assume a temperature of $T=3\times10^{5}$~K). This model
provides a good alternative description of the X-ray data, modelling both the
ionised edge and providing a good overall fit
$(\chi_{\nu}^{2}\approx1$). However both a large column of material 
($N_{H}\approx4.7\times10^{23}$ cm$^{-2}$) 
and a very high degree of ionisation 
($\xi=1.5\times10^{4}$~erg\ cm\ s$^{-1}$) are required
to reproduce the energy and depth of the edge feature. 

Extrapolation of the preferred hard X-ray spectral model 
to lower energies (0.7~keV) does not provide an adequate
description of the broad-band data. There remains a significant deficit of
counts present below $\sim$2~keV (see figure 4, panel 1), indicating
that significant intrinsic absorption (in excess of Galactic) is present. 
A neutral absorber ($N_{\rm H}\sim10^{22}\rm{cm}^{-2}$) gave a poor fit;
the spectrum is poorly modelled and an excess of counts is
present below 1 keV in the \asca\ data. However an
{\it ionised} absorber provides a good fit (figure 4, panel 2), with 
$N_{\rm H}\sim5\pm1.5\times10^{22}{\rm cm}^{-2}$ and $\xi\sim350\pm100$ (again assuming T\ $=3\times10^{5}$~K). 
Note that this material is not sufficiently ionised to 
reproduce the deep, ionised, iron K edge seen in the hard X-ray spectrum; 
the absorber effectively contributes a negligible amount 
($\tau<0.1$) to the iron K edge.
The absence of any appreciable optical
absorption (aside from Galactic) may also 
imply either a low intrinsic dust-to-gas
ratio, or that most of the dust close to the central engine 
has sublimated.

\begin{figure}
\begin{center}
\rotatebox{-90}{\includegraphics[width=5.5cm]{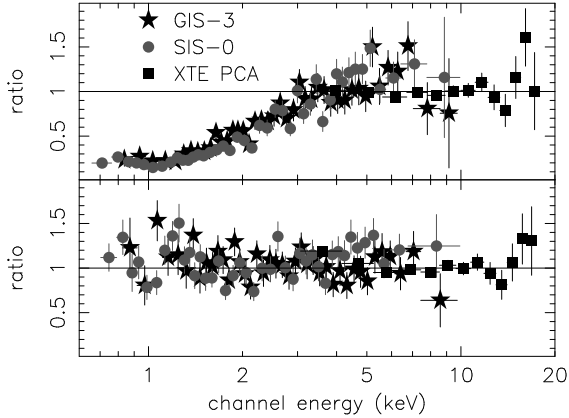}}
\end{center}
\caption{Data-ratio residuals for the broadband 0.7-18~keV
\asca\ and \xte\ spectrum of PDS~456. The first panel shows the
hard X-ray spectrum extrapolated down to lower energies. The
deficit of counts below 2 keV indicates the presence of substantial
soft X-ray absorption. The lower panel
shows the data residuals, after including a warm absorber
of column $\sim5\times10^{22}\rm{cm}^{-2}$.}  
\end{figure}

\section{ Temporal analysis }

Light curves were constructed by binning the data into 96 min orbital
bins, for the background-subtracted \xte\ (2--10~keV) and \asca\
(2--10 and 0.5--2~keV) data. 
The results are plotted in Figure~5.
The \xte\ light curve shows an increase by a factor of $\sim$2.1 in 3
orbits (17~ksec).
Note that the time-averaged \xte\ flux in the 2-10 keV band is
$ 8.5 \times 10^{-12} $ erg cm$^{-2}$ s$^{-1} $, and if this flare is real, 
it would be the most rapid variation ever seen in a radio-quiet quasar.
Unfortunately the (lower signal/noise) \asca\ observation had ended
before the flare began, so independent confirmation is not possible.
In order to assess the reality of this event, we consider two
possibilities that arise because \xte\ is a non-imaging instrument:
1) the flare could be due to an error in the background subtraction,
and 2) it could be due to variations in a contaminating source in the
\xte\ beam.

\begin{figure}
\begin{center}
\rotatebox{-90}{\includegraphics[width=5.5cm]{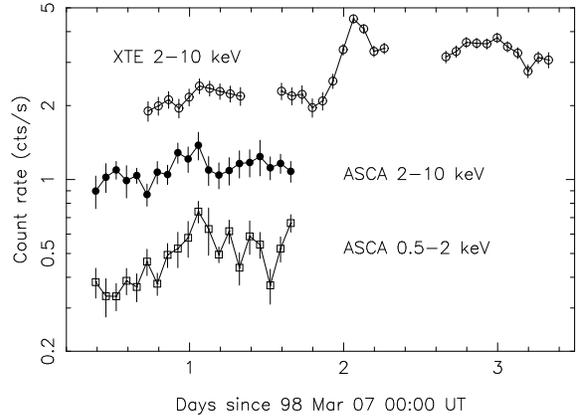}}
\end{center}
\caption{\xte\ and \asca\ light curves for PDS~456.
The \asca\ data were shifted upwards in this logarithmic plot by an 
arbitrary amount and the errors on the \xte\ data were increased by 
0.15 ct/s in quadrature to account for systematic errors.
The source appears to flare by a factor of $\sim$2.1 in 17~ksec,
corresponding to a light-crossing size of $ \sim 2 R_S $.}
\end{figure}

Figure~6 shows the derived on source data (2--10~keV L1)
along with two estimators of the background residuals (2--10~keV
Layer 2 [L2], and 20--40~keV Layer 1 [L1]).
The background model was determined in the same fashion for all three
light curves. As X-rays from this rather weak source should not
contribute significantly to the 2--10~keV L2 and 20--40~keV L1 count
rates, these residuals can be used to assess the quality of the 
background subtraction.
The 2--10~keV L2 data are well-behaved, consistent with the estimated
0.15~ct/s systematic error, but the 20--40~keV L1 show apparently small
periodic daily excursions on the $\sim$0.5~ct/s level. 
These latter excursions are not correlated with the flare, but instead
look like errors in modelling the daily effect of SAA passage.
Thus, we consider it unlikely that the
observed flare could be due entirely to poor background subtraction.

No contaminating sources of sufficient brightness were seen 
in the \asca\ image, but
that covered only $\sim$20\% of the $ \sim 1^o $ \xte\ field-of-view.
The RASSBSC (Voges \et 1996) was searched and no sources were found
within $ 1^o $ of PDS~456.
%However as \rosat\ has a much softer response, it is possible that a
%flat spectrum source could be excluded from the RASSBSC but detected
%by \xte.
\ginga\ background fluctuation data (Butcher \et 1997), which do cover
a similar energy band, suggest a $\ls$2\% probability of a
contaminating source at the 2~ct/s level; a similar probability of
2.5\% is found from the distribution of Piccinotti \et (1982).
However, there is no assurance that the \xte\ field-of-view does not
contain a highly variable contaminating source at a lower flux level.
For instance, the Butcher \et data suggest that a few sources could
be expected with mean \xte\ count rates of 0.2--0.4~ct/s.
If one of these is a narrow-line Seyfert~1 galaxy (NLS1s can show
factor of $\gs$5 in a few hours; e.g., Boller, Brandt \& Fink 1996),
or even a BL-Lac object,
this could produce a spurious flare identical to the one seen.
This is unlikely but cannot be ruled out.
Thus the most likely 
interpretation of the data is that the flare is intrinsic to PDS~456,
but other possibilities cannot be completely excluded. 
The discussion below assumes that the flare is intrinsic to the QSO.

\begin{figure}
\begin{center}
\rotatebox{-90}{\includegraphics[width=5.5cm]{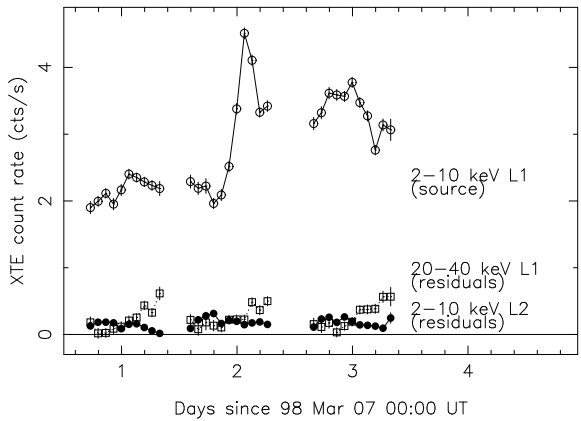}}
\end{center}
\caption{Background-subtracted on-source (2--10~keV L1) and residual
(2--10~keV L2 and 20--40~keV L1) light curves, plotted on a linear
scale, with no systematic errors included.}
\end{figure}

\section{ Discussion }

We find that PDS~456 has an unusual X-ray spectrum, with a
deep, highly ionised iron K edge and a broad (but poorly constrained) 
iron line, superimposed on a steep hard X-ray continuum. 
Lower energy residuals, corresponding to a warm absorber also seem 
to be present in the soft X-ray band. 
The high energy iron K features arise from a highly ionised reprocessor. One
possibility is through Compton reflection of hard X-rays off a
high ionisation accretion disk. An alternative model involves reprocessing
from line-of-sight matter (a `warm
absorber'). 
However, in the latter case, the degree of ionisation
required to produce such a deep and highly ionised edge 
seems rather high when compared to the warm absorbers observed in Seyfert 1s 
(Reynolds 1997); indeed it is possible that such a high ionisation absorber 
may not be thermally stable (see Reynolds \& Fabian 1995). 

Overall an ionised reflector appears a more physically appealing 
description of the hard X-ray data. 
Indeed strong, highly ionised, iron-K edge features are predicted 
by disk photoionisation models (e.g. Ross, Fabian \& Young 1999) and
are also thought to be observed in some high-state Galactic
black-hole candidates (e.g. Zycki, Done \& Smith 1997). The iron line and
edge produced in such highly ionised disks can be significantly
broadened through multiple Compton scatterings (e.g. Ross
\et 1999). However we were unable to differentiate here between a smeared
and a sharp iron edge, given the limited signal to noise 
of our observations.

The high ionisation of the reflector 
could imply a high accretion rate in PDS~456 (relative to the
Eddington limit), particularly as $\xi\propto\dot{m}^{3}$ in a
photoionised accretion disk (e.g. Matt, Fabian \& Ross 1993). 
This interpretation is consistent with the other X-ray properties of
PDS~456, namely a steep underlying continuum and rapid X-ray variability,
both of which are commonplace in NLS1s
(Vaughan \et 1999a and references therein). NLS1s are 
also thought to be accreting near the Eddington limit
(e.g. Pounds, Done \& Osborne 1995); indeed recent
evidence has been found in one NLS1 (Ark 564) for a spectrum
consistent with ionised disk reflection (Vaughan \et 1999b).  
Ionised iron line features have also been observed in other 
luminous radio-quiet quasars (Reeves \et 1997; Nandra \et 1997).
An unusually high accretion rate 
may even account for the complete lack of X-ray
spectral features in some of the most luminous quasars 
(Nandra \et 1995, Reeves \& Turner 1999). 
However PDS~456 seems to provide the clearest
example for the presence of ionised iron features in a high luminosity
AGN.

The one obvious flare in the X-ray light curve (figure 5) has a 
doubling time of $\sim$15~ks. 
This suggests, from simple light-crossing arguments, a maximum size of 
$l=4.5\times10^{12}$~m for the varying region. 
For a black hole of mass $10^{9}$M$_{\odot}$ (corresponding to 
PDS~456, with L$_{BOL}=10^{47}$ \erg, at the Eddington limit), this
implies that the X-ray flare occurs within a
region of less than 2 Schwarzschild radii (2R$_{S}$). (A more
conservative H$_{0}$ of 75 km s$^{-1}$ Mpc$^{-1}$ gives a region
of $<5R_{S}$, which is still tightly constrained). 
A smaller mass black hole would loosen this requirement somewhat, 
but would then imply a super-Eddington accretion rate. 
{\it Therefore one possible implication of the rapid
variability is accretion near to or greater than L$_{Edd}$.} The
variability also implies a (non-beamed) efficiency 
of converting matter to energy of $\sim$~5\%, 
close to the limit for a Schwarzschild black hole (see Fabian 1979).
Some Seyfert 1s (such as MCG-6-30-15, Reynolds \et 1995) can also exhibit rapid variability, limiting the X-ray emission to a few R$_{S}$.
Rapid flux changes have also been reported in 2 narrow-line QSOs; in PKS 0558-504, a 70\% increase is observed in 3 minutes (Remillard \et 1991) and PHL 1092 shows a reported increase by a factor of 3.8, using the \rosat\ HRI, in less than 5000~sec (Brandt \et 1999).
In any event, the flare in PDS~456 must come from a very compact region (for instance, a small hot spot on the disk),
presumably very close to the ``central engine" in which the luminosity
is actually generated.

In conclusion PDS~456 is a remarkable object, showing clear features of
a high ionisation reprocessor, one possible interpretation of which is through 
reflection off a highly ionised accretion disk. Both the high ionisation
spectral features and the extreme rapid variability suggest that the
super-massive black hole in PDS~456 could
be running at an unusually high accretion rate.

\bsp
\label{lastpage}
\end{document}